\documentstyle[12pt,a4wide]{article}
\newcommand{\be}{\begin{equation}}
\newcommand{\ee}{\end{equation}}
\newcommand{\bea}{\begin{eqnarray}}
\newcommand{\eea}{\end{eqnarray}}
\newcommand{\bean}{\begin{eqnarray*}}

\newcommand{\eean}{\end{eqnarray*}}
\font\upright=cmu10 scaled\magstep1 \font\sans=cmss10
\newcommand{\ssf}{\sans}
\newcommand{\stroke}{\vrule height8pt width0.4pt depth-0.1pt}
\newcommand{\Z}{\hbox{\upright\rlap{\ssf Z}\kern 2.7pt {\ssf Z}}}

\newcommand{\C}{{\rlap{\rlap{C}\kern 3.8pt\stroke}\phantom{C}}}
\newcommand{\R}{\hbox{\upright\rlap{I}\kern 1.7pt R}}
\newcommand{\PP}{\hbox{\upright\rlap{I}\kern 1.5pt P}}

\newcommand{\identity}{{\upright\rlap{1}\kern 2.0pt 1}}

\newcommand{\HH}{\mbox{\hbox{\upright\rlap{I}\kern 1.7pt H}}}

\newcommand{\fr}{\frac}

\newcommand{\pr}{\partial}
\newcommand{\hs}{\hspace{5mm}}

\newcommand{\dg}{\dagger}

\newcommand{\acc}{\\[3mm]}

\newcommand{\bi}{{DBI}}
\newcommand{\biga}{1+2(D_lU)^\dagger(D_lU)[1+(D_kU)^\dagger(D_kU)]
-(D_lU)^\dagger(D_kU)[(D_lU)^\dagger(D_kU)+(D_kU)^\dagger(D_lU)]}
\newcommand{\pp}{P_+}

\font\mybb=msbm10 at 11pt
\font\mybbb=msbm10 at 17pt
\def\bb#1{\hbox{\mybb#1}}
\def\bbb#1{\hbox{\mybbb#1}}

\def\bR {\bb{R}}

\def\bC {\bb{C}}

\def\bbC {\bbb{C}}
\newcommand{\CP}{\bC {\rm P}}
\newcommand{\CPP}{\bbC {\rm P}}

\input{epsf}

\newcommand{\news}{\setcounter{equation}{0}}
\begin{document}

\title{\vskip -70pt
\begin{flushright}
\end{flushright}\vskip 50pt
{\bf \large Non-BPS Dirac-Born-Infeld Solitons }\\[30pt]
\author{Theodora Ioannidou$^1$, George Papadopoulos$^2$ and
Paul M. Sutcliffe$^1$\\[10pt]
\\{\normalsize  {\sl 1.\, Institute of Mathematics,
University of Kent at Canterbury,}}\\ {\normalsize {\sl
Canterbury, CT2 7NZ, U.K.}}\\ {\normalsize{\sl Email :
T.Ioannidou@ukc.ac.uk}}\\ {\normalsize{\sl Email :
P.M.Sutcliffe@ukc.ac.uk}}\\
\\{\normalsize  {\sl 2.\, Department of Mathematics,
King's College London,}}\\ {\normalsize {\sl Strand,  London, NW10
2DU, U.K.}}\\ {\normalsize{\sl Email
: george.papadopoulos@kcl.ac.uk}}\\}}
\date{September 1999}
\maketitle

\begin{abstract}
\noindent We show that \CP$^n$  sigma model  solitons
solve the field equations of a Dirac-Born-Infeld (DBI) action
and, furthermore, we prove that the non-BPS soliton/anti-soliton solutions 
of the sigma model also solve the  DBI equations.
Using the moduli space approximation we compare 
 the dynamics of the BPS sigma model solitons with  that of the
 associated DBI solitons.
 We find that for
the \CP$^1$ case the metric on the moduli space of sigma model solitons is
identical to that of the moduli space of DBI solitons, but
for \CP$^n$ with $n>1$ we show
that the two metrics are not equal. We also consider the possibility of
similar non-BPS solitons in other DBI theories.

\end{abstract}

\newpage
\section{Introduction}
\news\ \ \ \ \ \
The low energy dynamics of branes in strings
and M-theory can be described by actions of  Dirac-Born-Infeld (DBI) type 
\cite{dbi}. The
various classical  solutions of these actions  admit a  geometric
interpretation in terms of brane configurations. In fact most of
these solutions have a bulk interpretation as intersecting
branes \cite{pktgpi}. The  part of the
DBI D-brane action which is quadratic in the
derivatives of the fields can be identified
 with  a Yang-Mills theory coupled to 
matter. This has led to a relation between intersecting
brane configurations and classical solutions of Yang-Mills
theories. One of the applications of this relation is that some
classical and quantum properties of  Yang-Mills theory can be understood
as geometric properties of  brane configurations \cite{witten}.

The fields of DBI type actions are scalars, vectors or tensors.
In certain cases, there is a consistent truncation of the DBI
action for which the vector and tensor fields are not
active\footnote{A field is active in a solution, if it  has a
non-trivial dependence on the brane worldvolume coordinates.}. The
remaining fields are scalars and the action reduces to a Dirac
one (or Nambu-Goto for the string case). Then powerful
geometrical methods can be used to investigate static
configurations, like those of (generalized) calibrations \cite{bbs, jan}. In
particular, bounds for the energy  can be established and the
solutions which saturate them can be found. The  part of this reduced
 DBI action which is quadratic in the derivatives of the scalars is that of
a non-linear sigma model.  So the possibility arises to interpret
various sigma model solitons in terms of brane configurations and
to relate the calibration bounds to sigma model ones. An example
was presented in \cite{BT, gpt}, where the sigma model lumps and
Q-kinks were understood in terms of M-2-brane configurations. In
particular it was observed that the sigma model energy bound
associated with lumps on a Taub-Nut target space is related to
the K\"ahler calibration  bound in the effective theory of the
M-2-brane in the presence of a Kaluza-Klein (KK)-monopole. Moreover,
 it was found
that the solutions that saturate the sigma model bound also solve
the corresponding bound of DBI. In this way the sigma model lumps
were embedded in the effective theory of the M-2-brane.

It is  expected  that  other sigma model solitons can be embedded
in a brane theory, although there are several restrictions on
the sigma model target space required by the consistency
of the background due to kappa symmetry.
Furthermore, for a DBI action it
is far from clear that solutions of the sigma model field
equations will also solve the DBI field equations since the former
is just an approximation of the latter. However, this appears to
be the case for a large class of solutions that saturate certain
sigma model energy bounds. This is probably due to supersymmetry
which protects them against higher derivative corrections. However,
as we shall demonstrate, there are sigma model solutions that do
not saturate a bound which are also  solutions of DBI
equations without any modification.

A large class of (2+1)-dimensional sigma models are those with
target spaces the complex projective spaces \CP$^n$. Since \CP$^n$
is a
 K\"ahler manifold, these sigma models
admit a supersymmetric extension with $N=2$ supersymmetry in three
dimensions ($N=1$ supersymmetry in four dimensions) and they are
described by chiral superfields.  The energy of the static sigma
model configurations is
bounded by the topological central charge of the supersymmetry
algebra.  The configurations that saturate this bound are solitons
which can be thought of as holomorphic curves in \CP$^n$. Many of the
properties of these solitons have been extensively  investigated,
 including their low energy scattering \cite{ward} as well as  soliton/anti-soliton
annihilation using numerical and other methods \cite{piette}. Another
novel class of solutions of \CP$^n$ sigma models are those describing
unstable soliton/anti-soliton states, which of course do not
saturate the above bound \cite{din, Za}.

In this paper, we consider a DBI action with target space 
$\CP^n$. Such a DBI action admits an energy bound
associated with a degree two K\"ahler calibration.
The part of the DBI action which is quadratic in the derivatives of the
fields can be identified with the $\CP^n$ sigma model.
 The calibrated submanifolds that
saturate the bound are holomorphic submanifolds of $\bR^2\times\CP^n$
and these are precisely the static solitons of the (2+1)-dimensional
 $\CP^n$ sigma model.
In addition, we shall show that
 the soliton/anti-soliton solutions of the \CP$^n$ 
sigma model are also solutions of the DBI action. These
solutions  saturate neither the sigma model nor  the
K\"ahler calibration bounds. 

Using the moduli space approximation, we compare the low energy dynamics of BPS
sigma model solitons with that of the associated DBI solitons
and we find that they may differ.
Finally, we investigate the possibility of other non-BPS solitons
in DBI theories.

\section{The \CPP$^n$ Sigma Model}
\news\ \ \ \ \ \
In this section we briefly review some properties of the
 (2+1)-dimensional   sigma model
with target space \CP$^n$ (for more details see eg. \cite{Za}).
For this, we  begin with the description a (2+1)-dimensional sigma
model with target space any K\"ahler manifold $M$ equipped with a
metric $h$ and compatible complex structure $K$. The sigma model
fields are maps $X$ from $\bR^{(1,2)}$ into $M$.
 The energy of a static sigma model
configuration ($X: \bR^2\rightarrow M$) is
\be
E_\sigma={1\over2} \int d^2x h_{IJ}\delta^{ij} \partial_iX^I
\partial_j X^J \ee where $i,j=1,2$ and $I,J=1,\dots, {\rm dim}M$.
Then \bea E_\sigma&=&{1\over 4} \int d^2x\, h_{IJ}\, \delta^{ij}\,
\big (\partial_i X^I\pm\epsilon^k{}_i K^I{}_L
\partial_k X^L\big)
\big (\partial_j X^J\pm\epsilon^\ell{}_j K^J{}_M \partial_\ell
X^M\big) \nonumber\\ &\mp&{1\over 2} \int\, d^2x\, (\omega_K)_{IJ}
\epsilon^{ij} \partial_i X^I\partial_j X^J\ , \eea where
$\epsilon$ is the Levi-Civita tensor  and $\omega_K$ is the
K\"ahler form of $K$. So we find
\be
E_\sigma\ge 2\pi |Q| \label{kahbound} \ee where
\be
Q={1\over 4\pi} \int d^2x (\omega_K)_{IJ} \epsilon^{ij} \partial_i
X^I\partial_j X^J 
\label{topq}\ee is the topological charge since $\omega_K$
is a closed form. Clearly, the configurations that saturate the
bound satisfy
\be
\partial_i X^I\pm\epsilon^k{}_i K^I{}_J \partial_k X^J=0\ .
\ee This is the Cauchy-Riemann equation and the solutions are
holomorphic curves in the K\"ahler manifold $M$.

For $M=\CP^n$, the sigma model fields $X$ can be
parameterized by a  complex $(n+1)$-component column vector $U$,
of unit length ie. $U^\dagger U=1.$ The sigma model Lagrangian is
given by
\be
{\cal L}_\sigma = (D_\mu U)^\dagger (D^\mu U) \label{lagsigma} \ee
where $D_\mu V=\partial_\mu V-(U^\dagger \partial_\mu U) V.$ Greek
indices run over the spacetime values 0,1,2 and we use the
Minkowski metric $\eta_{\mu\nu}=(-1,1,1).$ The sigma model
equation of motion which follows from (\ref{lagsigma}) is
\be
D_\mu D^\mu U+(D_\mu U)^\dagger (D^\mu U) U=0 \label{eomsigma} \ee
and all finite energy solutions are classified by the integer
valued topological charge (\ref{topq}), which in this parameterization is given
by
\be
Q=\frac{i}{2\pi} \int \ d^2x\, \epsilon^{jk} (D_j U)^\dagger (D_k
U) \ . \label{top} \ee The
configurations that saturate the bound satisfy
\be
D_j U=\mp i \epsilon^k{}_j D_k U. \label{bpssigma} \ee The BPS
soliton solutions are easily constructed as
\be
U=\frac{f}{\vert f\vert} \label{hol} \ee where $f$ is any vector whose
components are rational functions of the complex coordinate
$z=x_1+ix_2$. The topological charge $Q$ is positive in this case
and is equal to the highest degree of the rational functions which
occur in the entries of $f$. The anti-soliton solutions, which
have negative topological charge, are obtained in the same way but
with $f$ an anti-holomorphic function of $z$.

There are also non-BPS solutions, ie. solutions of the second
order equations (\ref{eomsigma}) which are not solutions of the
first order equations (\ref{bpssigma}), and therefore do not
saturate the energy bound \cite{din, Za}. All these solutions
can be obtained explicitly by making use of the operator $P_+$
acting on a general vector $g$ as
\be
P_+ g=\partial_z g-(g^\dagger \partial_z g)\frac{g}{\vert
g\vert^2}. \label{pplus} \ee The non-BPS solutions are then given
by a repeated application of this operator
\be
U=\frac{P_+^k f}{\vert P_+^k f \vert}, \ \ \ \ \ \ \ k=0,..,n
\label{nbpssolns} \ee starting with any initial holomorphic vector
$f.$ The operator $P_+$ can only be applied $n$ times, since after
this number of applications the original holomorphic vector is
converted into an anti-holomorphic vector and $P_+$ annihilates 
anti-holomorphic vectors. Thus this operator
converts BPS solitons into BPS anti-solitons, but intermediate
solutions with $0<k<n$ correspond to non-BPS soliton/anti-soliton
configurations. Note that in the case of the \CP$^1$ model the
operator can be applied only once and converts BPS soliton
solutions to BPS anti-soliton solutions, so there are no non-BPS
solutions in this case.

\section{The \CPP$^n$ DBI Model}
\news\ \ \ \ \ \
To define the \CP$^n$ DBI model we begin with the spacetime
$\bR^{(1,2)}\times \CP^n$ with metric
\be d\tilde s^2=  ds^2(\bR^{(1,2)})+ ds^2(\CP^n)\ 
\label{metr} \ee
where 
 $ds^2(\CP^n)$ is the
Fubini-Study metric on \CP$^n.$
We then consider a (2+1)-dimensional embedded submanifold on which
the action is obtained as the worldvolume of the induced metric obtained
by pulling back the spacetime metric (\ref{metr}) to the embedded submanifold.
Explicitly,
 \be
I_{DBI}=\int d^3x \sqrt{{\rm
det}(\tilde \gamma_{\mu\nu})}\label{lagbi} \ee
where $\tilde \gamma_{\mu\nu}$ is the induced metric ie. the pull
back of $d\tilde s^2$ to the submanifold. We choose the static gauge,
in which the spacetime coordinates of the submanifold are identified
with the $\bR^{(1,2)}$ part of the spacetime metric, and then clearly
the part of the 
action (\ref{lagbi}) which is quadratic in the derivatives of the scalars
 can be identified with the action of the \CP$^n$ sigma model 
described in the previous section. In the above DBI action we have
set the Born-Infeld field to zero, which is a consistent truncation.

As in the case of the \CP$^n$ sigma models in the previous section, we
shall seek static solutions of this system. 
The energy\footnote{This energy functional
arises naturally in the calibration bound and includes the vaccum energy.}
 of such solutions is
\be
E_{DBI}= \int d^2x
\sqrt{{\rm det}(\tilde\gamma_{ij})}, \ee where
$\tilde\gamma_{ij}$ is the pull back of the spatial part of the
metric (\ref{metr}). The solutions that minimize the energy are
minimal surfaces in $\bR^2\times \CP^n$. Since
$\bR^2\times \CP^n$ is a K\"ahler manifold,  it is well known that
there is a bound for the energy which is described by a K\"ahler
calibration. In particular
\be
E_{DBI}\ge 2\pi |Q| \label{bound} \ee where $Q$ is a topological charge induced
by pulling back the K\"ahler form
\be
\omega=  dx^1\wedge dx^2+  \omega_K \ee 
onto the embedded submanifold, and 
 $\omega_K$ is the K\"ahler form on \CP$^n$. 
The
configurations that saturate the bound are holomorphic curves in
$\bR^2\times \CP^n$. So we find that the
 configurations that saturate the
energy bound of the sigma model also saturate  the calibration
bound for the DBI solitons.

In terms of the parameterization, as a unit length column vector $U$,
 introduced in the previous section for the sigma model, the DBI 
Lagrangian becomes  
\be
{\cal L}_{DBI}=\sqrt{-\mbox{det}(\eta_{\mu\nu}+(D_\mu U)^\dagger
(D_\nu U) +(D_\nu U)^\dagger (D_\mu U))}-1\  \label{lagbi2} \ee 
and the static energy is
\be
E_{DBI}= \int d^2x
\sqrt{{\rm det}(\eta_{ij}+(D_i U)^\dagger
(D_j U) +(D_j U)^\dagger (D_i U))}-1 \label{dbien}\ee 
where we have now subtracted off the vacuum energy.

In this formulation it is an easy exercise to verify the
energy bound (\ref{bound}), where $E_{DBI}$ and $Q$ are given
by (\ref{dbien}) and (\ref{top}) respectively, and to show that
the sigma model solitons, given by (\ref{hol}) with $f$ a holomorphic
(or anti-holomorphic) function, saturate this bound.

\section{Non-BPS DBI Solitons}
\news\ \ \ \ \ \
We shall now find that, remarkably, the non-BPS sigma model solitons
 are also solutions of the
DBI field equations.  To prove this we begin by computing
the  DBI field equations for static configurations that follow
from the variation of (\ref{lagbi2}). 
After a little algebra
we find that the static field equations are
\small
\bea
D_i\{\frac{(D_iU)[1+2(D_jU)^\dagger(D_jU)]-(D_jU)[(D_iU)^\dagger(D_jU)+
(D_jU)^\dagger(D_iU)]}{\sqrt{\biga}}\}\nonumber\\
=-\frac{U\{(D_iU)^\dagger(D_iU)[1+2(D_jU)^\dagger(D_jU)]
-(D_iU)^\dagger(D_jU)[(D_iU)^\dagger(D_jU)+(D_jU)^\dagger(D_iU)]\}.}
{\sqrt{\biga}}\nonumber\\ \ \label{eombi} \eea \normalsize It is
convenient to introduce the notation $T_{ij}\equiv
(D_iU)^\dagger(D_jU)$ and rewrite equation (\ref{eombi}) in the
form \bea
D_1\{\frac{(D_1U)(1+2T_{22})-(D_2U)(T_{12}+T_{21})}{\sqrt{
1+2T_{11}+2T_{22}+4T_{11}T_{22}-(T_{12}+T_{21})^2}}\} +\nonumber\\
D_2\{\frac{(D_2U)(1+2T_{11})-(D_1U)(T_{12}+T_{21})}{\sqrt{
1+2T_{11}+2T_{22}+4T_{11}T_{22}-(T_{12}+T_{21})^2}}\} +\nonumber\\
\frac{U[T_{11}+T_{22}+4T_{11}T_{22}-(T_{12}+T_{21})^2]}{\sqrt{
1+2T_{11}+2T_{22}+4T_{11}T_{22}-(T_{12}+T_{21})^2}}=0.\nonumber\\
\ \label{eom2bi} \eea Next we observe that if we impose the
following constraints
\be
T_{11}=T_{22}, \ \ T_{12}=-T_{21} \label{con} \ee then equations
(\ref{eom2bi}) can be simplified to
\be
D_1 D_1 U+D_2 D_2 U + U(T_{11}+T_{22})=0 \ee which are exactly the
 sigma model field equations for static configurations (\ref{eomsigma}).
Thus if we can find static solutions of the sigma model equations
which also satisfy the constraints (\ref{con}) then they will also
be static solutions of the \bi\ equations. The BPS solitons solve
the equations (\ref{bpssigma}) and so automatically obey the
constraints (\ref{con})
 (and have the additional property that $T_{12}=\pm iT_{11}$). We shall
now show that the sigma model non-BPS solitons also satisfy the
constraints.

In terms of the complex variable $z=x_1+ix_2$ the pair of real
equations (\ref{con}) can be written as the single complex
equation
\be
(D_zU)^\dagger(D_{\bar z}U)=0. \label{ccon} \ee From the
definition of $P_+$, given by (\ref{pplus}), the following
properties may be proved (see eg. \cite{Za}) when $f$ is a
holomorphic vector,
\begin{eqnarray}
\label{bbb} &&(\pp^k f)^\dg \,\pp^l f=0, \hs \hs \hs k\neq l\acc
&&\pr_{\bar{z}}\left(\pp^k f\right)=-\pp^{k-1} f \fr{|\pp^k
f|^2}{|\pp^{k-1} f|^2}, \hs \pr_{z}\left(\fr{\pp^{k-1}
f}{|\pp^{k-1} f|^2}\right)=\fr{\pp^k f}{|\pp^{k-1}f|^2}. \hs
\label{aaa}
\end{eqnarray}
Using these properties it is elementary to show that if $U$ is a
non-BPS solution given by (\ref{nbpssolns}) then
\be
D_zU=\frac{P_+^{k+1}f}{\vert P_+^k f\vert}, \ \ \ \ D_{\bar
z}U=-P_+^{k-1}f\frac{\vert P_+^k f\vert}{\vert P_+^{k-1}f\vert^2}.
\ \ \ee 
Hence equation (\ref{ccon}) follows immediately from the
orthogonality property (\ref{bbb}). This completes the proof that
the non-BPS solitons are also solutions of the \bi\ model.

These non-BPS solutions have the interpretation of
 soliton/anti-soliton states just as in the sigma model. 
As in the sigma model \cite{Za} these solutions are therefore
expected to be unstable configurations. The global
properties of such solutions as submanifolds of $\CP^n$ have been
investigated in \cite{wood}. 

The reader may wonder if it is possible to see why the non-BPS
solitons solve the \bi\ equations directly from the Lagrangian. Of
course, in general it is inconsistent  to substitute constraints
in  the Lagrangian since the critical points of the constrained
system are generally not critical points of the full theory. For
BPS solitons this is not a concern because they saturate the
energy bound and so are automatically critical points. It is
therefore enough to show that the \bi\ Lagrangian reduces to the
sigma model Lagrangian when the BPS constraints are imposed.
However, non-BPS solutions require a little more thought.

As we want to investigate when the \bi\ Lagrangian (or
equivalently energy density since we are dealing with the static
sector)
 reduces to that of the sigma model
then we compute that \bea & &(1+{\cal E}_\sigma)^2-(1+{\cal
E}_{DBI})^2\nonumber\\
&=&(1+T_{11}+T_{22})^2-(1+2T_{11}+2T_{22}+4T_{11}T_{22}-
(T_{12}+T_{21})^2)\nonumber\\
&=&(T_{11}-T_{22})^2+(T_{12}+T_{21})^2. \label{diff} \eea In the
final line we recognize the constraints (\ref{con}) and there are
two crucial points to be noted. The \bi\ energy density is
identical to that of the sigma model if and only if the
constraints (\ref{con}) hold. The first point is that these
constraints are weaker than the BPS equations and have more
solutions, in particular we have shown that the non-BPS solitons
satisfy the constraints. The second point is that the above
expression is quadratic in the constraints, so its variation is
proportional to the constraints and vanishes after their
imposition. This explains why the critical points of the
constrained system, which is the sigma model, are also critical
points of the full \bi\ theory.

\section{Sigma Model vs DBI Dynamics}
\news\ \ \ \ \ \
\label{dynamics} As we have seen, 
static sigma model BPS
solitons also solving the  DBI
field equations. It is therefore natural to compare
the dynamics of slowly moving BPS solitons in the sigma model and
in the \bi\ model, using the moduli space approximation \cite{Ma1}. Naively
it might be expected that the low energy dynamics of BPS solutions
in the two theories will be the same since they have the same low
energy limit. However, the agreement of the two Lagrangians is to
quadratic order in all the derivatives of the fields ie. space
and time, whereas the moduli space approximation assumes that time
derivatives are small but there is no truncation of the spatial
derivatives. Thus it is not clear whether the dynamics of 
 the \bi\ solitons  will match that of 
the sigma model solitons. The moduli space of sigma model
solitons is a K\"ahler manifold \cite{Ru2}.

The moduli space approximation truncates the full field dynamics
to motion on the BPS soliton moduli space. Applying the BPS
soliton equation (\ref{bpssigma}) the \bi\ Lagrangian
(\ref{lagbi}) can be written as
\be
{\cal
L}_{DBI}=\sqrt{(1+2T_{11})(1+2T_{11}-2T_{00}-4(T_{11}T_{00}-T_{01}T_{10}))}-1
\ee 
where the notation is as in section four. Expanding
out the square root and neglecting terms which are higher order
than quadratic in the {\sl time} derivatives, we obtain, after the
integration over space
\be
L_{DBI}=2\pi\vert Q\vert -\int d^2x\big( T_{00} +
2(T_{11}T_{00}-T_{01}T_{10})\big) \ . \ee 
The first term is just
the potential energy of a charge $Q$ soliton and the remaining
term defines a metric on the $Q$-soliton moduli space, with
respect to which the dynamics is approximated by geodesic motion.
The first term in the integrand is the kinetic energy of the sigma
model, thus the metric on the moduli space of DBI solitons will
be equal to that of the sigma model solitons if and only if the term (which
we now write out in full)
\be
K\equiv 2\int \{(D_1U)^\dagger(D_1U)(D_0U)^\dagger(D_0U)-
(D_0U)^\dagger(D_1U)(D_1U)^\dagger(D_0U)\} \ d^2x \label{k} 
\ee
vanishes or is equal to a total time derivative.

For the \CP$^1$ model this term is indeed zero. The easiest way to
see this is to choose the parameterization (which can be done
without loss of generality using the local $U(1)$ symmetry)
\be
U=\frac{1}{\sqrt{1+\vert w\vert^2}}\pmatrix{1\cr w} 
\ee 
in terms of which it is easily found that
\be
D_\mu U =\frac{\partial_\mu w}{(1+\vert
w\vert^2)^{3/2}}\pmatrix{-\bar w \cr 1}. 
\ee 
Thus $D_0 U$ is
proportional to $D_1 U$ and hence the integrand in (\ref{k}) is
identically zero. However, this cancellation is a unique property
of the \CP$^1$ case and derives from the fact that it has only one
(complex) field.

To demonstrate that in the \CP$^n$ model with $n>1$ the term
(\ref{k}) can be non-zero it is enough to consider the \CP$^2$
case (since \CP$^2$ is a totally geodesic submanifold of \CP$^n$
for $n>2$). Writing the \CP$^2$ field as
\be
U=\frac{1}{\sqrt{1+\vert w_1\vert^2 +\vert w_2\vert^2}}
\pmatrix{1\cr w_1\cr w_2} \ee then
\be
K=\int \frac{2 \vert \partial_0 w_1\partial_z w_2-\partial_0
w_2\partial_z w_1\vert^2}{ (1+\vert w_1\vert^2 +\vert
w_2\vert^2)^3} \ d^2x. \label{kcp2} 
\ee 
If either $w_1$ or $w_2$
are identically zero then this term vanishes, corresponding to the
embedding of \CP$^1$ inside \CP$^2$, but generally (\ref{kcp2}) is
non-zero. Finally, we need to check that this term is not a total
time derivative, otherwise it would not contribute to the geodesic
equations. The easiest way to show this is by a simple example.

An axially symmetric \CP$^2$ soliton of 
charge $Q$ has the form
\be
w_1=\alpha z^Q-\beta, \ \ \ w_2=\alpha z^Q+\beta 
\ee 
where $\alpha$ and $\beta$ are complex parameters related to the size
and shape of the soliton \cite{PRZ}. Taking $\alpha$ and $\beta$
to be time dependent the integrand in (\ref{kcp2}) is axially
symmetric and the integration is elementary to arrive at
\be
K=\frac{2\pi Q \vert
\partial_0\beta\vert^2}{(1+2\vert\beta\vert^2)^2} 
\ee 
which is clearly not a total time derivative.

One expects that in a supersymmetric extension the  $\CP^n$ DBI solitons
for $n>1$ break more
supersymmetry\footnote{We remark that the corresponding
sigma model solitons always break half of the supersymmetry of the
$N=2$ (2+1)-dimensional sigma model.} than those in $\CP^1$.
Because of this, the above observation regarding the low energy dynamics
 of solitons,  is
related to a similar observation in \cite{GP} that the small
perturbations around supersymmetric solitons that preserve
less than $1/4$ of the maximal spacetime supersymmetry in the
Maxwell Theory-Sigma Model approximation of the DBI  action do not
solve the  perturbed  DBI field equations. However if the solutions
preserve $1/4$ of  spacetime supersymmetry, then the
perturbations of the Maxwell Theory-Sigma Model approximation also
solve the perturbed  DBI field equations. This indicates that for
BPS solitons which preserve enough supersymmetry,  supersymmetry
protects the sigma model soliton moduli metric from higher derivative
corrections.

In summary we have shown that for \CP$^1$ the slow motion dynamics
of DBI solitons is identical to that of the 
sigma model solitons, but for
$n>1$ the dynamics of slowly moving  \CP$^n$ DBI solitons is
different from that of the sigma model solitons. This is despite 
the fact that the
low energy truncation of the \bi\ Lagrangian gives 
the sigma model one and that
the BPS solutions of the two systems are the same.

\section{Non-BPS Born-Infeld Field Configurations}
\news\ \ \ \ \ \
The investigation of the relation between BPS and non-BPS
solutions of DBI theory and those of sigma models, that we have described,
 can be
extended to include the relation between BPS and non-BPS
solutions of DBI theory and those of Yang-Mills coupled to matter systems.
Recently $SU(n)$ monopole solutions have been constructed which
solve the second order Yang-Mills-Higgs equations but are not
solutions of the first order Bogomolny equations. Not only are
these solutions three-dimensional analogues of the non-BPS
\CP$^{n-1}$ sigma model solutions but in fact the sigma model
 solutions are
used explicitly to obtain the monopoles \cite{IS3}. An obvious
candidate in the search for non-BPS \bi\ solitons is therefore to
consider a system of $n$ parallel D3-branes in type IIB string
theory, since its low energy truncation reproduces the
Yang-Mills-Higgs Lagrangian. However, we shall make a simple
observation that suggests that the non-Bogomolny Yang-Mills-Higgs
monopoles will not be solutions of the \bi\ equations in this
case. This in fact may not be a surprise. For example, in \cite{gp} it has
been observed that even some BPS solutions of Yang-Mills theory
do not solve the Born-Infeld field equations.

Although the effective action of a single D3-brane in type IIB
string theory is described by an abelian \bi\ Lagrangian a system
of $n$ parallel D3-branes is expected to be described by a $U(n)$
\bi\ theory, the most promising of which is the proposal of
Tseytlin \cite{Ts}. In the static case, with one active adjoint
scalar, the energy density of the D3-brane worldvolume theory can
be conveniently written as \cite{Ha}
\be
{\cal E}_{BI}=\mbox{STr}(\sqrt{\mbox{det}(\delta_{ab}+F_{ab})}-1)
\label{d3en} \ee where $a,b$ go from 1 to 4 and a dimensional
reduction is performed in the $x_4$ direction with the usual
identification of the scalar field as $\Phi=A_4$. The gauge group
is $SU(n)$ (an overall $U(1)$ factor decouples as the centre of
mass) and STr denotes the trace over gauge indices of the weighted
sum over all permutations of the non-commutative products
\cite{Ts}. This is required in order to make sense of the ordering
ambiguities involved in computing the determinant.

Expanding (\ref{d3en}) to quadratic order in the fields reproduces
the static Yang-Mills-Higgs energy density
\be
{\cal E}_{YMH}=\mbox{Tr}(\frac{1}{2}D_i\Phi
D_i\Phi+\frac{1}{4}F_{jk}F_{jk}) \label{ymhen} \ee where
$i=1,2,3.$ The BPS monopole solutions of (\ref{ymhen}), which
satisfy the Bogomolny equation
\be
D_i\Phi=\pm\frac{1}{2}\epsilon_{ijk}F_{jk} \label{bog} \ee are
also solutions of the Born-Infeld theory (\ref{d3en}), which can
be shown by noting that the two energies (\ref{d3en}) and
(\ref{ymhen}) are equal upon substitution of the Bogomolny
equation \cite{Ha,Br}. The charge $k$ monopole solution describes
$k$ D-strings stretched between the $n$ D3-branes and this can be
seen explicitly by graphing the eigenvalues of the scalar field
$\Phi$ over $\bR^3$ \cite{Ha}.

In order to determine whether the non-Bogomolny monopole solutions
\cite{IS3} of (\ref{ymhen}) are solutions of the Born-Infeld
theory we follow the procedure given in (\ref{diff}) for the sigma
model case, and compute when the Yang-Mills-Higgs and Dirac-Born-Infeld
energies agree. Ignoring the trace operation and treating the
matrices as if they were abelian (which can be justified using the
symmetrized trace \cite{Ha,Br}) we compute that \bea 2[(1+{\cal
E}_{YMH})^2-(1+{\cal E}_{BI})^2]= (D_2\Phi
D_3\Phi+F_{12}F_{13})^2+ (D_1\Phi
D_3\Phi+F_{21}F_{23})^2\nonumber\\ +(D_1\Phi
D_2\Phi+F_{31}F_{32})^2 +(D_2\Phi F_{12}+D_3\Phi F_{13})^2
+(D_1\Phi F_{21}+D_3\Phi F_{23})^2\nonumber\\ +(D_1\Phi
F_{31}+D_2\Phi F_{32})^2 +\frac{1}{2}((D_1\Phi)^2-(F_{23})^2)^2
+\frac{1}{2}((D_2\Phi)^2-(F_{13})^2)^2\nonumber\\
+\frac{1}{2}((D_3\Phi)^2-(F_{12})^2)^2. \ \hskip 8cm 
\eea 
As in
the sigma model case, (\ref{diff}), we find that the result can
again be written as a sum of squares, but this time there is an
important difference in that the constraints (under which the two
energies agree) now contain the Bogomolny equations (\ref{bog})
explicitly (see the last three terms). Thus in this case, in
contrast to the sigma model example, the constraints are solved
only by the Bogomolny monopoles. Although this does not prove that
the non-Bogomolny monopoles do not solve the Born-Infeld theory,
it strongly suggests that they do not and certainly shows that the
feature of non-BPS solitons in the sigma model does not apply in
the same way to this example.

\section {Concluding Remarks}
\news\ \ \ \ \ \
We have shown that the $\CP^n$ sigma model  BPS solitons 
also solve the field equations of a Dirac-Born-Infeld
action. Furthermore, we have shown that certain non-BPS $\CP^n$ sigma model
 solutions, which correspond to soliton/anti-soliton configurations, are also
solutions of the Dirac-Born-Infeld action.
We have also investigated the dynamics of the $\CP^n$ DBI
solitons and  found that they do not  coincide with the sigma model
dynamics unless $n=1$.
Finally, we explored the possibility of similar non-BPS solutions
in Yang-Mills theories.

The possible D-brane interpretation of our results remains an open problem.
Although it is expected that some sigma model solitons can be embedded
 in a brane theory there are several restrictions on the DBI action,
 such as kappa symmetry. In particular, this implies that the sigma
model target space should be a solution of the supergravity field equations.
Since there are no supergravity solutions which are topologically
$\CP^n$ then it is not obvious that any of our solutions have a D-brane
interpretation. However, there are several possibilities which may
admit a D-brane interpretation. For example, the near horizon geometry
of the M-2-brane is $AdS_4\times S^7$ and since $S^7$ is a circle
bundle over $\CP^3$ then by a Kaluza-Klein reduction along the circle
fibre it is possible to obtain a background which includes $\CP^3 \cite{DLP}.$
Our sigma model solitons can be embedded as solutions in this case but
unfortunately this requires a singular embedding and so there is no
obvious D-brane interpretation for this example.

\section*{Acknowledgements}
\news\ \ \ \ \ \
Many thanks to Gary Gibbons,  Nick Manton and Wojtek Zakrzewski for useful
discussions. PMS acknowledges the EPSRC for an Advanced Fellowship
and the grant GR/L88320. GP is supported by a Royal Society Research
Fellowship.\\


\newpage



\begin{thebibliography}{99}

\bibitem{dbi} J.H. Hughes, J. Liu and J. Polchinski, {\sl Supermembranes},
Phys. Lett. {\bf B180} (1986) 370.\\
 R.G. Leigh, {\sl Dirac-Born-Infeld action
from Dirichlet Sigma Model}, Mod. Phys. Lett. {\bf A4} (1989) 2767 .
\\ E. Bergshoeff, E. Sezgin and
P.K. Townsend, {\sl Supermembranes and 11 dimensional
supergravity}, Phys. Lett. {\bf B189} (1987) 75. \\ P.S. Howe and
E. Sezgin, {\sl D=11, P=5}, Phys. Lett. {\bf B394} (1997) 62,
hep-th/9611008.

\bibitem{pktgpi} G. Papadopoulos and P.K. Townsend,
{\sl Intersecting M-branes}, Phys. Lett.  {\bf B380} (1996) 273,
hep-th/9603087.

\bibitem{witten} E. Witten, {\sl Solutions of four-dimensional field
theories via M-theory}, Nucl. Phys. {\bf B500} (1997) 3,
hep-th/9703166.

\bibitem{bbs}K. Becker, M. Becker and A. Strominger, {\sl Fivebranes, 
membranes and
non-perturbative string theory}, Nucl. Phys. {\bf B456} (1995)
130.

\bibitem{jan}J. Gutowski and G. Papadopoulos, {\sl AdS Calibrations},
hep-th/9902034.

\bibitem{BT} E. Bergshoeff and P.K. Townsend, {\sl Solitons
on the supermembrane}, JHEP {\bf 05} (1999) 021, hep-th/9904020.

\bibitem{gpt} J. Gutowski, G. Papadopoulos and P.K. Townsend,
{\sl Supersymmetry and Generalized Calibrations}, hep-th/9905156.

\bibitem{ward}R.S. Ward, {\sl Slowly-Moving lumps in the $\CP^1$ model in (2+1)
dimensions}, Phys. Lett. {\bf B158} (1985) 424.\\
R. Leese, {\sl Low-energy scattering of solitons in the $\CP^1$ model}, Nucl. Phys.
{\bf B344} (1990) 33. 

\bibitem{piette} B. Piette, P.M. Sutcliffe and W.J. Zakrzewski, 
{\sl Soliton Antisoliton scattering in 2+1 dimensions}, Int. J. of Mod.
Phys. {\bf 3} (1992) 637.

\bibitem{din} A.M. Din and W.J. Zakrzewski, {\sl General classical solutions of
the $\CP^{n-1}$ model}, Nucl. Phys. {\bf B174} (1980) 397.

\bibitem{Za} W. J. Zakrzewski, {\it Low dimensional sigma models} (IOP, 1989).

\bibitem{wood} J. Eells and J.C. Wood, {\sl Harmonic maps from surfaces to
complex projective spaces}, Advances in Mathematics {\bf 49} (1983) 217.
 
\bibitem{Ma1} N.S. Manton, {\sl A remark on the scattering of BPS
monopoles},  Phys. Lett. {\bf B110} (1982) 54.

\bibitem{Ru2} P.J. Ruback, {\sl  Sigma model solitons and their moduli space
metric}, Commun. Math. Phys. {\bf 116} (1988) 645.

\bibitem{PRZ} B.M.A.G. Piette, M.S.S. Rashid and W.J.
Zakrzewski, {\sl Soliton scattering in the $\CP^2$ model},
 Nonlinearity {\bf 6} (1993) 1077.

\bibitem{GP} J. Gutowski and G. Papadopoulos, {\sl The moduli spaces
of worldvolume solitons}, Phys. Lett. {\bf B432} (1998) 97,
hep-th/9811207; {\sl The dynamics of D-3-brane dyons and toric
hyper-K\"ahler manifolds}, Nucl. Phys. {\bf B551} (1999) 650,
hep-th/9811207.

\bibitem{IS3} T. Ioannidou and P.M. Sutcliffe, {\sl Non-Bogomolny SU(N) BPS Monopoles}, 
hep-th/9905169.

\bibitem{gp} G.W. Gibbons and G. Papadopoulos, {\sl Calibrations and
Intersecting Branes}, Commun. Math. Phys. {\bf 202} (1999) 593,
hep-th/9803163.

\bibitem{Ts} A.A. Tseytlin, {\sl On non-abelian generalization of Born-Infeld
action in string theory},  Nucl. Phys. {\bf B501}  (1997) 41, hep-th/9701125.

\bibitem{Ha} A. Hashimoto, {\sl The shape of branes pulled by strings},
Phys. Rev. {\bf D57} (1998) 6441, hep-th/9711097.

\bibitem{Br} D. Brecher, {\sl BPS states of the non-abelian Born-Infeld
actions},  Phys. Lett. {\bf B442}  (1998) 117, hep-th/9804180.

\bibitem{DLP} M.J. Duff, H. L\"u and C.N. Pope, {\sl $AdS_5\times S^5$
Untwisted}, Nucl. Phys. {\bf B532} (1998) 181, hep-th/9803061;\\
 M.J. Duff, {\sl Anti-de Sitter space, branes, singletons,
superconformal field theories and all that},
Int. J. Mod. Phys. {\bf A14} (1999) 815, hep-th/9808100.


\end{thebibliography}
\end{document}